\documentclass[journal=jmcmar,manuscript=article,layout=twocolumn]{achemso}
\setkeys{acs}{
    abbreviations = false,
    keywords = false,
}

\usepackage[version=3]{mhchem} 
\usepackage{siunitx}
\usepackage[super]{nth}
\usepackage{caption}
\usepackage{subcaption}
\usepackage{multirow,tabularx}
\usepackage{adjustbox}
\usepackage{textgreek}
\usepackage{dblfloatfix}

\DeclareSIUnit{\calorie}{cal}
\DeclareSIUnit{\Calorie}{\kilo\calorie}

\let\oldmaketitle\maketitle
\let\maketitle\relax
\author{Parham Rezaee}
\affiliation{Department of Biophysics, School of Biological Sciences, Tarbiat Modares University, Tehran, Iran}
\alsoaffiliation{UNESCO-UNISA-iTLABS Africa Chair in Nanoscience and Nanotechnology (U2ACN2), College of Graduate Studies, University of South Africa (UNISA), Pretoria, South Africa}
\author{Shahab Rezaee}
\affiliation{Department of Biophysics, School of Biological Sciences, Tarbiat Modares University, Tehran, Iran}
\author{Malik Maaza}
\affiliation{UNESCO-UNISA-iTLABS Africa Chair in Nanoscience and Nanotechnology (U2ACN2), College of Graduate Studies, University of South Africa (UNISA), Pretoria, South Africa}
\author{Seyed Shahriar Arab}
\affiliation{Department of Pediatrics, University of California, San Diego, La Jolla, CA 92093, USA}
\email{ssarab@health.ucsd.edu}

\title[]
  {Screening of BindingDB database ligands against EGFR, HER2, Estrogen, Progesterone and NF-$\kappa$B receptors based on machine learning and molecular docking}


\begin{document}

\begin{tocentry}

\includegraphics[width=\textwidth]{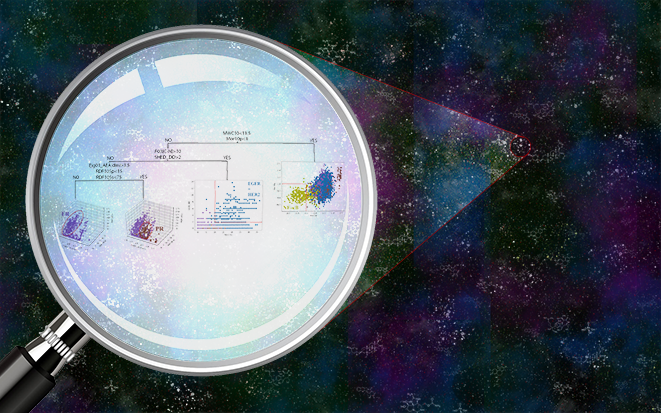}

\end{tocentry}

\twocolumn[
\begin{@twocolumnfalse}
\oldmaketitle
\begin{abstract}
Breast cancer, the second most prevalent cancer among women worldwide, necessitates the exploration of novel therapeutic approaches. To target the four subgroups of breast cancer ``hormone receptor-positive and HER2-negative, hormone receptor-positive and HER2-positive, hormone receptor-negative and HER2-positive, and hormone receptor-negative and HER2-negative'' it is crucial to inhibit specific targets such as EGFR, HER2, ER, NF-$\kappa$B, and PR.

In this study, we evaluated various methods for binary and multiclass classification. Among them, the GA-SVM-SVM:GA-SVM-SVM model was selected with an accuracy of 0.74, an F1-score of 0.73, and an AUC of 0.94 for virtual screening of ligands from the BindingDB database. This model successfully identified 4454, 803, 438, and 378 ligands with over 90\% precision in both active/inactive and target prediction for the classes of EGFR+HER2, ER, NF-$\kappa$B, and PR, respectively, from the BindingDB database. Based on to the selected ligands, we created a dendrogram that categorizes different ligands based on their targets. This dendrogram aims to facilitate the exploration of chemical space for various therapeutic targets.

Ligands that surpassed a 90\% threshold in the product of activity probability and correct target selection probability were chosen for further investigation using molecular docking. The binding energy range for these ligands against their respective targets was calculated to be between -15 and -5 \si{\kilo\calorie\per\mole}. Finally, based on general and common rules in medicinal chemistry, we selected 2, 3, 3, and 8 new ligands with high priority for further studies in the EGFR+HER2, ER, NF-$\kappa$B, and PR classes, respectively.

\end{abstract}
\end{@twocolumnfalse}
]
\section{Introduction}
Breast cancer, characterized by the highest mortality rate among various cancer types, is a widespread condition. The development and progression of breast cancer are facilitated by the interaction of estrogen and progesterone receptors with breast cells\cite{ismail_insilico_2018}. These hormones, estrogen and progesterone, bind to their respective receptors in the cytoplasm, leading to dimerization and subsequent entry into the nucleus. Additionally, they bind to estrogen and progesterone response elements located near the promoters of target genes. In a study by Shirazi, it was demonstrated that estradiol, an estrogen hormone, alone promoted the growth of MCF-7 cells compared to the control group. Conversely, tamoxifen, a well-known estrogen blocker, arrested the proliferation of MCF-7 cells for a minimum of five days. When both the stimulator (estradiol) and blocker (tamoxifen) were applied together, the level of stimulation in MCF-7 cell growth was reduced. Consequently, both estrogen and progesterone contribute to the initiation and acceleration of breast cancer. An excessive presence of these receptors indicates hormone (ER+/PR+) mediated breast cancer. Targeting estrogen and progesterone hormones can aid in the identification of potent inhibitors for breast cancer\cite{zhou_retracted_2014, gnanaselvan_structure-based_2024}.

\begin{figure*}[!ht]
    \centering
    \includegraphics[width=0.98\textwidth]{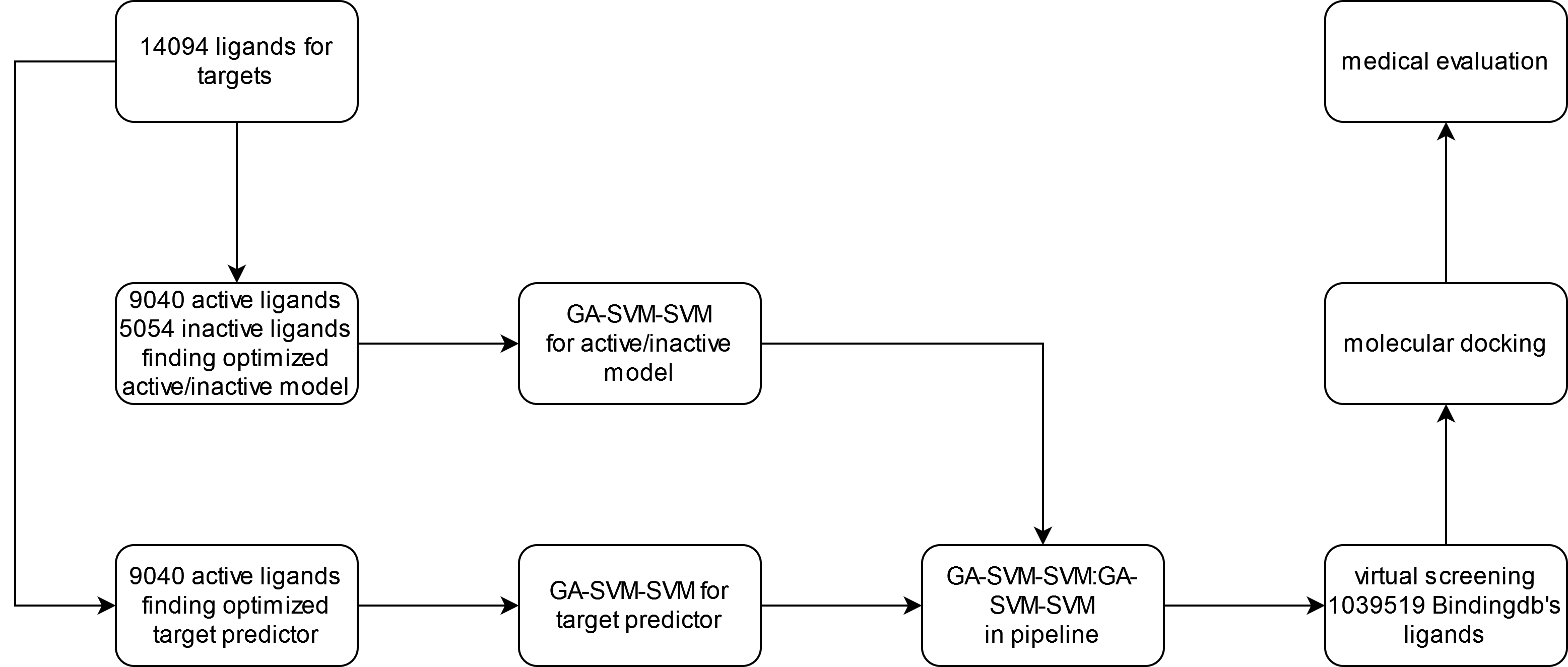}
    \caption{The flowchart of these procedures.}
    \label{fig:flowchart}
\end{figure*}

EGFR is a transmembrane glycoprotein with a crucial role in cell signaling. It consists of a ligand-binding domain and a tyrosine kinase domain. Binding with endogenous epidermal growth factor (EGF) leads to dimerization, autophosphorylation, and activation of downstream pathways, resulting in cell proliferation and differentiation\cite{schwartz_protein_2011}. EGFR is a major target in breast cancer, and anti-EGFR agents have shown efficacy, especially in patients with specific EGFR mutations. Tyrosine kinase inhibitors (TKIs) like gefitinib, erlotinib, and lapatinib have been used to inhibit EGFR overexpression\cite{yang_afatinib_2015, johnston_s.r.d._lapatinib:_2006}.

HER2 is a protein with tyrosine kinase activity encoded by the neu/erbB2/c-erbB2 oncogene. It belongs to the EGFR family and plays a crucial role in the development of normal and malignant breast tissue. Approximately 30\% of human breast carcinomas exhibit HER2 amplification\cite{brennan_her2/neu:_2000}. HER2 interacts with insulin-like growth factor receptor-1 and estrogen receptor, initiating cell signaling. Targeting HER2 has been a focus in anti-cancer drug development, with the discovery of inhibitors like neratinib and afatinib. However, further trials are needed to confirm their efficacy against breast cancer. HER2 overexpression can increase the activity of MMP-2 and MMP-9 proteases, promoting invasiveness of breast cancer cells\cite{li_suppression_2014}. It also amplifies VEGF expression and MMP-9 activity, potentially triggering angiogenic responses. Clinical studies have shown that breast cancer cases with Erbb2 gene amplification have reduced responsiveness to certain treatments compared to cases with normal ErbB2 expression\cite{purawarga_matada_molecular_2022,asadollahi-baboli_shuffling_2013}.

Nuclear factor-kappa B (NF-$\kappa$B) is a transcription factor that regulates the expression of genes involved in cell proliferation, immunological responses, and inflammation. It contributes to the development of breast tumors, lymphoma, and colorectal cancer\cite{zubair_role_2013}. In breast cancer, NF-$\kappa$B activation occurs downstream of EGFR signaling, specifically in the ER-negative subtype. HER-2 overexpression leads to the activation of the PI3K/Akt pathway and induction of NF-$\kappa$B. NF-$\kappa$B plays a role in angiogenesis by stimulating the expression of VEGF and IL-8. It can activate two signaling pathways: the classical (canonical) pathway and the alternative (noncanonical) pathway\cite{alberti_ligand-dependent_2012}. Drugs like lapatinib and certain microtubule disruptors activate NF-$\kappa$B. In vitro studies have shown that ginseng inhibits COX-2 and NF-$\kappa$B activation in breast cancer cell lines\cite{peralta_american_2009}.

Despite remarkable advancements in the field of basic life sciences and biotechnology, the process of drug discovery and development (DDD) continues to be slow and costly. On average, it takes around 15 years and approximately US\$2 billion to develop a small-molecule drug\cite{sun_why_2022}. While clinical studies are widely acknowledged as the most expensive phase in drug development, the greatest potential for time and cost savings lies in the earlier stages of discovery and preclinical research. Preclinical efforts alone account for over 43\% of pharmaceutical expenses, in addition to significant public funding\cite{sun_why_2022,alamdari_monte_2012}. This is primarily due to the high attrition rate observed at every step, ranging from target selection and hit identification to lead optimization and the selection of clinical candidates. Furthermore, the substantial failure rate in clinical trials, currently at 90\%\cite{bajorath_computer-aided_2015}, can largely be attributed to issues originating in the early stages of discovery, such as inadequate target validation or suboptimal properties of ligands. Discovering faster and more accessible methods to identify a broader range of high-quality chemical probes, hits, and leads with optimal absorption, distribution, metabolism, excretion, and toxicology (ADMET) as well as pharmacokinetics (PK) profiles during the early phases of DDD would significantly enhance outcomes in preclinical and clinical studies. Consequently, this would enable the development of more effective, accessible, and safer drugs\cite{sadybekov_computational_2023,jalali-heravi_qsar_2011}.

In this study, first, we developed two models with the purpose of identifying active/inactive molecules and determining the target of each ligand. Subsequently, we constructed a pipeline using these models. This pipeline was utilized to screen the bindingDB ligands, employing various thresholds of model precision. The selected ligands were subjected to molecular docking to assess their binding energy with their respective targets. Additionally, we applied several established principles in medicinal chemistry to prioritize further investigation of these selected molecules, such as molecular dynamics, in vitro and in vivo studies. Furthermore, we examined the significance of the features employed in creating the target predictor model, aiming to identify a simple rule for acceptable accurate target recognition as a common rule. The procedure's flowchart of this study is illustrated in Figure \ref{fig:flowchart}.

\section{Materials and methods}
In this study, our dataset consisted of inhibitors targeting various breast cancer targets: 7341 for EGFR, 2182 for HER2, 1859 for ER, 1273 for NF-$\kappa$B, and 1439 for PR. To obtain these compounds, we downloaded five specific sdf files from the Binding database\cite{gilson_bindingdb_2016} website, each file corresponding to a particular class of inhibitors. These sdf files were then converted to gjf files using the OpenBabel\cite{oboyle_open_2011} software. The 3D structures of all the molecules were optimized using the Austin model 1 Hamiltonian implemented in Gaussian software\cite{frisch_gaussian_2009}. After optimization, the molecules were used to calculate molecular descriptors with the help of Alvadesc\cite{roy_alvadesc:_2020} software. A total of 5668 descriptors, including 0-, 1-, 2-, and 3D descriptors, were generated. To streamline the dataset, descriptors with constant values in 90\% of the compounds were removed. Additionally, among descriptors with a correlation above 0.9, the one exhibiting higher pair correlation with all other descriptors was kept and the others were automatically excluded. Following these processes, 1461 descriptors remained for further analysis.

\begin{table}[!b]
    \caption{Number of active and inactive molecules for each class}
    \label{tab:num_active_inactive}
    \begin{tabular}{|l|c|c|c|}
        \hline
        \textbf{Target} & \textbf{active} & \textbf{inactive} & \textbf{total} \\ \hline
        EGFR            & 4922            & 2419              & \textbf{7341}  \\ \hline
        ER      & 1223            & 636               & \textbf{1859}  \\ \hline
        HER2            & 1393            & 789               & \textbf{2182}  \\ \hline
        NF-$\kappa$B    & 447             & 826               & \textbf{1273}  \\ \hline
        PR              & 1055            & 384               & \textbf{1439}  \\ \hline
        \textbf{total}  & \textbf{9040}   & \textbf{5054}     & \textbf{14094} \\ \hline
    \end{tabular}
\end{table}

Each sdf file contains activity information pertaining to a specific molecule, indicating the affinity of that molecule towards different therapeutic targets. We extracted the activity information from the downloaded sdf files for each class of molecules, and saved it in separate vectors. The enumeration of the collected data can be found in Table \ref{tab:num_active_inactive}. Molecules with $IC_{50}$, $K_i$, and $EC_{50}$ values below 2000 nM were categorized as active inhibitors, while those with values exceeding 2000 nM were considered inactive. Also we removed molecules with activity more than 10000 nM as outlier data from the inactive dataset. Both active and inactive molecules were utilized to develop and evaluate the active/inactive classifiers. These models serve the purpose of screening extensive databases and identifying new potent molecules for the treatment of breast cancer.

A dataset of 1039519 molecules was gathered from the BindingDB database. The same preparation process used for the breast cancer inhibitors described earlier was applied to these downloaded molecules. Additionally, the descriptors selected for the breast cancer inhibitors were also chosen for these molecules. This resulted in a data matrix of size $1039519 \times 1461$, which was used for further analysis.

We employed various methods, including k-best, K-Nearest Neighbors (KNN), Gaussian Naive Bayes (GNB), Quadratic Discriminant Analysis (QDA), Random Forest (RF), and Support Vector Machine (SVM), to independently select descriptors for the active/inactive and target classifiers. To optimize the selection of descriptors across all data, we utilized a Genetic Algorithm (GA) in an optimal manner. The GA started with a population size of 200 and evolved through a maximum of 1000 generations, employing a crossover rate of 0.5 and a mutation rate of 0.2. The estimator was configured with the aforementioned methods, utilizing 5-fold cross-validation and an accuracy scoring function. The only difference between the feature selection processes of the active/inactive and target classifiers was the maximum number of features. The active/inactive classifier allowed a maximum of 64 features, while the target classifier allowed a maximum of 128 features.

We utilized the chosen features to create an optimized binary classifier for predicting active/inactive molecules. Various methods, including K-nearest neighbors (KNN), support vector machine (SVM), decision tree (DT), random forest (RF), naive Bayes (NB), linear discriminant analysis (LDA), and quadratic discriminant analysis (QDA), were employed for this purpose. To maximize the performance of each method, we conducted a grid search to identify the best parameters for constructing the model. Since the data reality is not imbalanced and due to a lack of studies on active molecules, we selected 461 active and 461 inactive molecules from the dataset to create a balanced training dataset for binary classification.

For the target classifier, we employed methods such as KNN, SVM, DT, logistic regression (LR), RF, NB, Gaussian naive Bayes (GNB), LDA, and QDA. Similar to the binary classifier, we utilized grid search to identify the optimal parameters for constructing the model. To ensure balance in the training dataset, we selected 132 active molecules for each class for the multiple classifier as well.

\begin{table*}[!ht]
    \caption{The evaluation of GA-SVM-SVM and GA-RF-RF for binary and GA-SVM-SVM and GA-QDA-SVM models for therapeutic classification.}
    \label{tab:classifier_selection}
    \begin{tabular}{|l|c|c|c|c|}
        \hline
        \textbf{GA-SVM-SVM}   & \textbf{precision} & \textbf{recall} & \textbf{f1-score} & \textbf{support} \\ \hline
        \textbf{active}       & 0.78               & 0.70            & 0.74              & 461              \\ \hline
        \textbf{inactive}     & 0.73               & 0.80            & 0.76              & 461              \\ \hline
        \textbf{GA-RF-RF}     & \textbf{precision} & \textbf{recall} & \textbf{f1-score} & \textbf{support} \\ \hline
        \textbf{active}       & 0.75               & 0.73            & 0.74              & 461              \\ \hline
        \textbf{inactive}     & 0.74               & 0.76            & 0.75              & 461              \\ \hline
        \textbf{GA-SVM-SVM}   & \textbf{precision} & \textbf{recall} & \textbf{f1-score} & \textbf{support} \\ \hline
        \textbf{EGFR+HER2}    & 0.95               & 0.92            & 0.93              & 132              \\ \hline
        \textbf{ER}           & 0.91               & 0.95            & 0.93              & 132              \\ \hline
        \textbf{NF-$\kappa$B} & 0.94               & 0.96            & 0.95              & 132              \\ \hline
        \textbf{PR}           & 0.96               & 0.93            & 0.95              & 132              \\ \hline
        \textbf{GA-QDA-SVM}   & \textbf{precision} & \textbf{recall} & \textbf{f1-score} & \textbf{support} \\ \hline
        \textbf{EGFR+HER2}    & 0.94               & 0.92            & 0.93              & 132              \\ \hline
        \textbf{ER}           & 0.95               & 0.95            & 0.95              & 132              \\ \hline
        \textbf{NF-$\kappa$B} & 0.93               & 0.94            & 0.93              & 132              \\ \hline
        \textbf{PR}           & 0.96               & 0.96            & 0.96              & 132              \\ \hline
    \end{tabular}
\end{table*}

Based on the selected best models for the active/inactive and target classifiers, they were combined into a single pipeline as a decision model for predicting the activity and target of each BindingDB ligand. The decision-making process for the model's predictions of activity and target was constrained by a certainty threshold of 0.8, 0.85, and 0.9 for the active/inactive classifier, and 0.9 for the target classifier.

Autodock Vina\cite{trott_autodock_2010} was utilized for molecular docking to calculate the binding affinities between ligands and their respective targets. The grid box resolution was set with specific coordinates for each target: (EGFR)(PDB ID: 1M17) had coordinates of 23.424, 1.310, 51.002 along the x, y, and z axes, respectively, with a grid spacing of 0.2 \si{\angstrom}; (HER2)(PDB ID: 3PP0) had coordinates of 17.563, 16.689, 26.321; (PR)(PDB ID: 1A28) had coordinates of 17.038, 0.145, 74.798; (ER)(PDB ID: 2IOK) had coordinates of 19.050, 35.696, 52.244; and (NF-$\kappa$B)(PDB ID: 4KIK) had coordinates of 48.268, 31.589, -57.885. These coordinates were used to define the binding sites for the docking process. The grid dimensions were set at $25.2 \times 25.2 \times 25.2$ \si{\angstrom}. The control ligands were initially docked with the binding sites of the five receptors, and the resulting interactions were compared with standard reference ligands.

To prioritize new ligands for further studies, such as molecular dynamics and others, we utilized various rules such as Lipinski, Pfizer, GSK, and golden triangle rules. Additionally, important parameters for drug production, including QED, SAscore, and MCE-18, were calculated using ADMETlab 2.0\cite{xiong_admetlab_2021}.

\begin{table*}[!ht]
    \caption{The evaluation of GA-SVM-SVM:GA-SVM-SVM, GA-RF-RF:GA-QDA-SVM, GA-SVM-SVM:GA-QDA-SVM, and GA-RF-RF:GA-SVM-SVM models in the pipeline.}
    \label{tab:pipeline}
    \begin{tabular}{|l|ccc|c|}
        \hline
        \textbf{\begin{tabular}[c]{@{}l@{}}GA-SVM-SVM:\\ GA-SVM-SVM\end{tabular}} & \multicolumn{1}{c|}{\textbf{precision}} & \multicolumn{1}{c|}{\textbf{recall}} & \textbf{f1-score} & \textbf{support} \\ \hline
        \textbf{N/A}                                                              & \multicolumn{1}{c|}{0.75}               & \multicolumn{1}{c|}{0.74}            & 0.74              & 461              \\ \hline
        \textbf{EGFR+HER2}                                                        & \multicolumn{1}{c|}{0.68}               & \multicolumn{1}{c|}{0.74}            & 0.71              & 99               \\ \hline
        \textbf{ER}                                                             & \multicolumn{1}{c|}{0.83}               & \multicolumn{1}{c|}{0.83}            & 0.83              & 115              \\ \hline
        \textbf{NF-$\kappa$B}                                                            & \multicolumn{1}{c|}{0.61}               & \multicolumn{1}{c|}{0.61}            & 0.61              & 123              \\ \hline
        \textbf{PR}                                                               & \multicolumn{1}{c|}{0.78}               & \multicolumn{1}{c|}{0.79}            & 0.78              & 124              \\ \hline
        \textbf{accuracy}                                                         & \multicolumn{3}{c|}{0.74}                                                                          & 922              \\ \hline
        \textbf{macro avg}                                                        & \multicolumn{1}{c|}{0.73}               & \multicolumn{1}{c|}{0.74}            & 0.73              & 922              \\ \hline
        \textbf{weighted avg}                                                     & \multicolumn{1}{c|}{0.74}               & \multicolumn{1}{c|}{0.74}            & 0.74              & 922              \\ \hline
        \textbf{\begin{tabular}[c]{@{}l@{}}GA-RF-RF:\\ GA-QDA-SVM\end{tabular}}   & \multicolumn{1}{c|}{\textbf{precision}} & \multicolumn{1}{c|}{\textbf{recall}} & \textbf{f1-score} & \textbf{support} \\ \hline
        \textbf{N/A}                                                              & \multicolumn{1}{c|}{0.73}               & \multicolumn{1}{c|}{0.76}            & 0.74              & 461              \\ \hline
        \textbf{EGFR+HER2}                                                        & \multicolumn{1}{c|}{0.68}               & \multicolumn{1}{c|}{0.70}            & 0.69              & 99               \\ \hline
        \textbf{ER}                                                             & \multicolumn{1}{c|}{0.82}               & \multicolumn{1}{c|}{0.83}            & 0.83              & 115              \\ \hline
        \textbf{NF-$\kappa$B}                                                            & \multicolumn{1}{c|}{0.56}               & \multicolumn{1}{c|}{0.50}            & 0.53              & 123              \\ \hline
        \textbf{PR}                                                               & \multicolumn{1}{c|}{0.80}               & \multicolumn{1}{c|}{0.76}            & 0.78              & 124              \\ \hline
        \textbf{accuracy}                                                         & \multicolumn{3}{c|}{0.73}                                                                          & 922              \\ \hline
        \textbf{macro avg}                                                        & \multicolumn{1}{c|}{0.72}               & \multicolumn{1}{c|}{0.71}            & 0.71              & 922              \\ \hline
        \textbf{weighted avg}                                                     & \multicolumn{1}{c|}{0.72}               & \multicolumn{1}{c|}{0.73}            & 0.73              & 922              \\ \hline
        \textbf{\begin{tabular}[c]{@{}l@{}}GA-SVM-SVM:\\ GA-QDA-SVM\end{tabular}} & \multicolumn{1}{c|}{\textbf{precision}} & \multicolumn{1}{c|}{\textbf{recall}} & \textbf{f1-score} & \textbf{support} \\ \hline
        \textbf{N/A}                                                              & \multicolumn{1}{c|}{0.75}               & \multicolumn{1}{c|}{0.74}            & 0.74              & 461              \\ \hline
        \textbf{EGFR+HER2}                                                        & \multicolumn{1}{c|}{0.68}               & \multicolumn{1}{c|}{0.73}            & 0.70              & 99               \\ \hline
        \textbf{ER}                                                             & \multicolumn{1}{c|}{0.82}               & \multicolumn{1}{c|}{0.84}            & 0.83              & 115              \\ \hline
        \textbf{NF-$\kappa$B}                                                            & \multicolumn{1}{c|}{0.60}               & \multicolumn{1}{c|}{0.59}            & 0.60              & 123              \\ \hline
        \textbf{PR}                                                               & \multicolumn{1}{c|}{0.78}               & \multicolumn{1}{c|}{0.78}            & 0.78              & 124              \\ \hline
        \textbf{accuracy}                                                         & \multicolumn{3}{c|}{0.74}                                                                          & 922              \\ \hline
        \textbf{macro avg}                                                        & \multicolumn{1}{c|}{0.73}               & \multicolumn{1}{c|}{0.74}            & 0.73              & 922              \\ \hline
        \textbf{weighted avg}                                                     & \multicolumn{1}{c|}{0.74}               & \multicolumn{1}{c|}{0.74}            & 0.74              & 922              \\ \hline
        \textbf{\begin{tabular}[c]{@{}l@{}}GA-RF-RF:\\ GA-SVM-SVM\end{tabular}}   & \multicolumn{1}{c|}{\textbf{precision}} & \multicolumn{1}{c|}{\textbf{recall}} & \textbf{f1-score} & \textbf{support} \\ \hline
        \textbf{N/A}                                                              & \multicolumn{1}{c|}{0.73}               & \multicolumn{1}{c|}{0.76}            & 0.74              & 461              \\ \hline
        \textbf{EGFR+HER2}                                                        & \multicolumn{1}{c|}{0.69}               & \multicolumn{1}{c|}{0.71}            & 0.70              & 99               \\ \hline
        \textbf{ER}                                                             & \multicolumn{1}{c|}{0.82}               & \multicolumn{1}{c|}{0.82}            & 0.82              & 115              \\ \hline
        \textbf{NF-$\kappa$B}                                                            & \multicolumn{1}{c|}{0.57}               & \multicolumn{1}{c|}{0.51}            & 0.54              & 123              \\ \hline
        \textbf{PR}                                                               & \multicolumn{1}{c|}{0.80}               & \multicolumn{1}{c|}{0.77}            & 0.78              & 124              \\ \hline
        \textbf{accuracy}                                                         & \multicolumn{3}{c|}{0.73}                                                                          & 922              \\ \hline
        \textbf{macro avg}                                                        & \multicolumn{1}{c|}{0.72}               & \multicolumn{1}{c|}{0.71}            & 0.72              & 922              \\ \hline
        \textbf{weighted avg}                                                     & \multicolumn{1}{c|}{0.73}               & \multicolumn{1}{c|}{0.73}            & 0.73              & 922              \\ \hline
    \end{tabular}
\end{table*}

\section{Results and discussion}
EGFR and HER2 receptors, shows 83.71\% similarity in their residues using sequence alignment with BLOSUM weight matrix and have a large similarity in their 3D structure using Needleman-Wunsch alignment algorithm with BLOSUM-62 similarity matrix (Fig. S1 and S2). Moreover, near 70\% of ligands in BindingDB database with EGFR and HER2 targets, were identical. According to these reasons, we merged two classes of EGFR and HER2 to just one EGFR/HER2 class.

The active/inactive and target classifiers utilized 128 and 64 selected features, respectively (see tables S1 and S2). Various models were created using the aforementioned methods. The top two models for the active/inactive classifier were GA-SVM-SVM and GA-RF-RF, while for the target classifier, they were GA-SVM-SVM and GA-QDA-SVM. The GA-SVM-SVM binary classifier was constructed using the radial basis function (RBF) kernel with a gamma value of 0.1 and a regularization term of 1. Similarly, GA-RF-RF was built using the Gini function with 400 trees and maximum depth until all leaves were pure. The forest construction involved the use of bootstrap sampling. Furthermore, GA-SVM-SVM and GA-QDA-SVM multi-classifiers were employed with the RBF kernel, a gamma value of 0.01, and a regularization term of 10. The classification was performed using the one-vs-one strategy, which has been shown to provide higher prediction accuracy compared to the one-vs-rest approach\cite{jalaliheravi_integrated_2013}. As shown in Table \ref{tab:classifier_selection}, the active/inactive classifiers GA-SVM-SVM and GA-RF-RF achieved precision, recall, and F1-scores all above 0.7. Likewise, the target classifiers GA-SVM-SVM and GA-RF-RF achieved precision, recall, and F1-scores all above 0.9.

\begin{table*}[!b]
    \caption{Number of selected BindingDB molecules for each targets according to the threshold of 0, 80, 85, and 90\% decision certainity for active/inactive prediction and the threshold of 0 and 90\% decision certainity for target prediction.}
    \label{tab:virtual_screening_result}
    \begin{tabular}{|l|ccccc|}
        \hline
        \multicolumn{1}{|c|}{} & \multicolumn{5}{c|}{\textbf{Threshold}}                                                                                                                             \\ \hline
        \textbf{classes}       & \multicolumn{1}{c|}{\textbf{0:0}} & \multicolumn{1}{c|}{\textbf{0:90}} & \multicolumn{1}{c|}{\textbf{80:90}} & \multicolumn{1}{c|}{\textbf{85:90}} & \textbf{90:90} \\ \hline
        \textbf{EGFR+HER2}     & \multicolumn{1}{c|}{172498}       & \multicolumn{1}{c|}{95123}         & \multicolumn{1}{c|}{19796}          & \multicolumn{1}{c|}{11068}          & 4454           \\ \hline
        \textbf{ER}    & \multicolumn{1}{c|}{54101}        & \multicolumn{1}{c|}{22876}         & \multicolumn{1}{c|}{3613}           & \multicolumn{1}{c|}{2029}           & 803            \\ \hline
        \textbf{NF-$\kappa$B}  & \multicolumn{1}{c|}{45452}        & \multicolumn{1}{c|}{16400}         & \multicolumn{1}{c|}{2499}           & \multicolumn{1}{c|}{1257}           & 438            \\ \hline
        \textbf{PR}            & \multicolumn{1}{c|}{67323}        & \multicolumn{1}{c|}{14109}         & \multicolumn{1}{c|}{2300}           & \multicolumn{1}{c|}{1116}           & 378            \\ \hline
    \end{tabular}
\end{table*}

\begin{figure}[!ht]
    \centering
    \includegraphics[width=0.98\columnwidth]{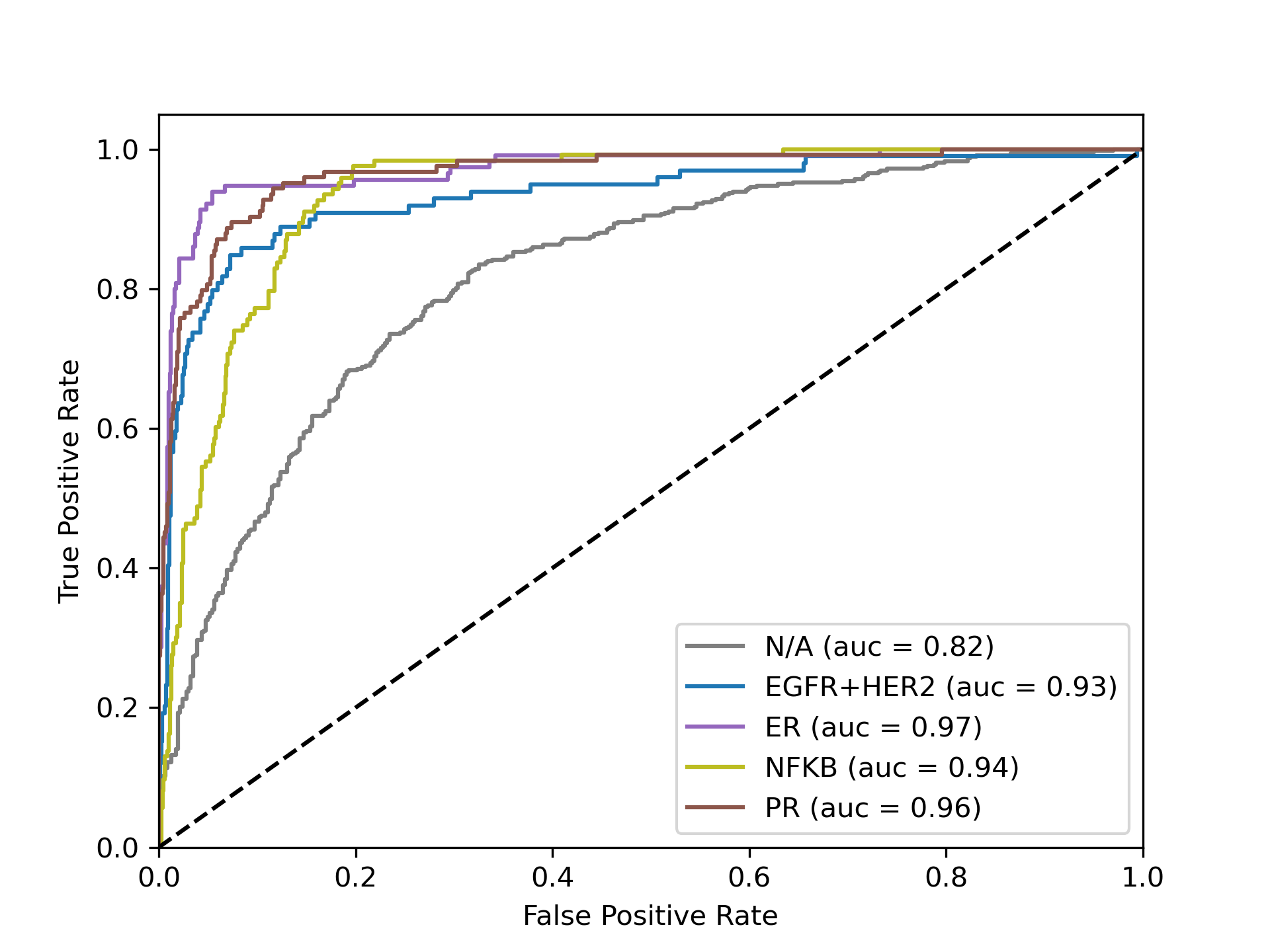}
    \caption{The ROC plots for different classes with one-vs-rest strategy.}
    \label{fig:roc}
\end{figure}

\begin{figure*}[!ht]
    \centering
    \includegraphics[width=0.98\textwidth]{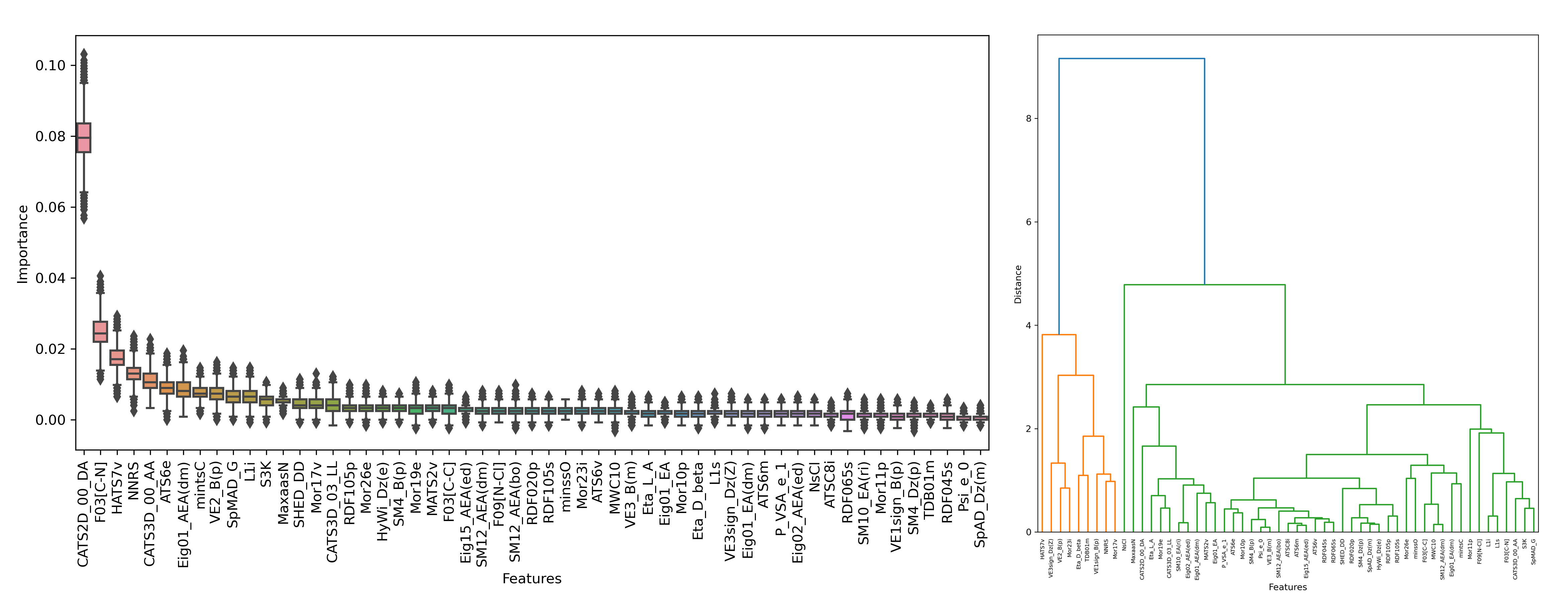}
    \caption{The left plot shows the importance of features using permutation importance method using GA-SVM-SVM model for target prediction and the right one demonstrates hierarchical clustering dendrogram using pearson method to find the correlation distance of each features.}
    \label{fig:features}
\end{figure*}

\begin{figure*}[!ht]
    \centering
    \includegraphics[width=0.98\textwidth]{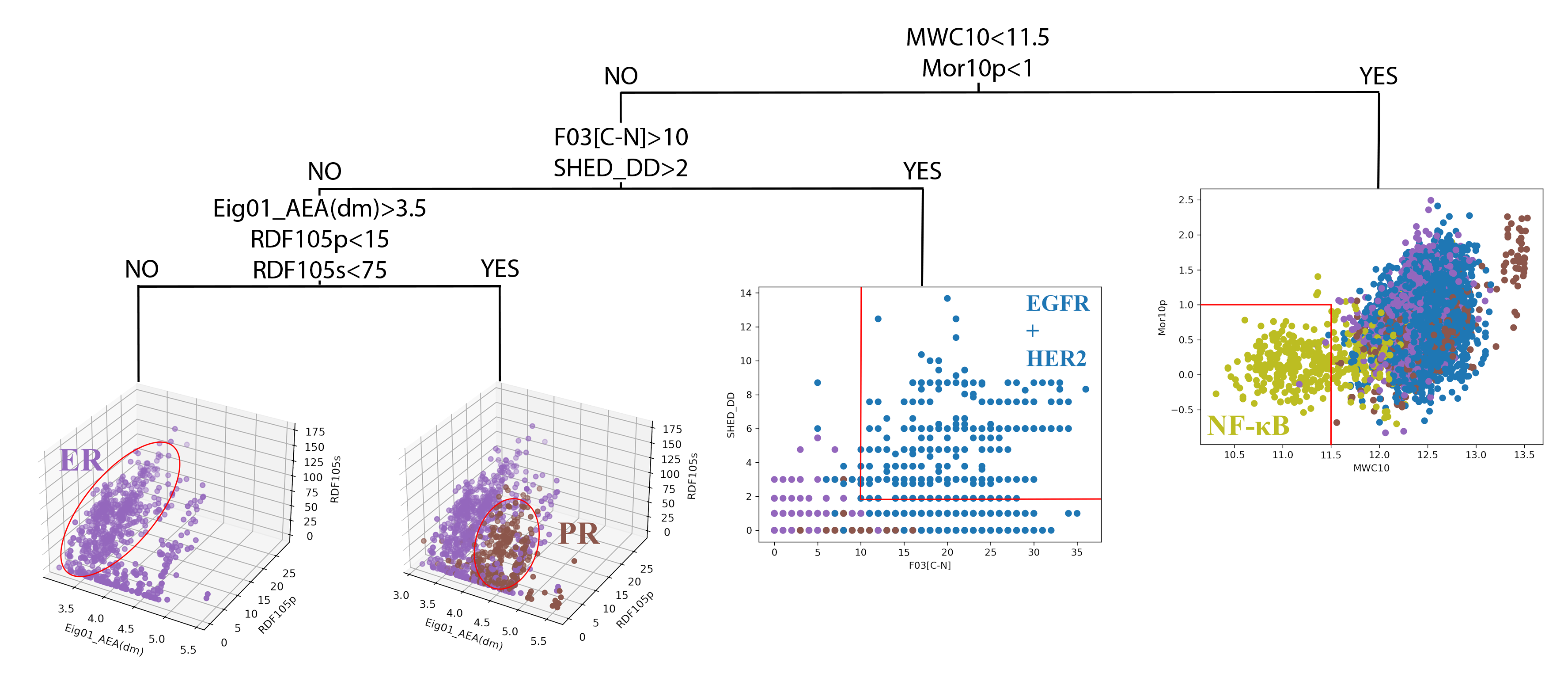}
    \caption{A simple questionnaire dendrogram to separate ligands with a number of features and determine the targets of them.}
    \label{fig:questionnaire_dendrogram}
\end{figure*}

Subsequently, we generated a pipeline by combining different permutations of the selected model. The precision, recall, f1-score, and support of these models are presented in Table \ref{tab:pipeline}. Among the options, the GA-SVM-SVM:GA-SVM-SVM model emerged as the most suitable pipeline, displaying superior performance compared to others. This approach achieved an accuracy of 0.74 and an AUC of 0.94. Figure \ref{fig:roc} showcases the ROC plots for each class using the one-vs-rest strategy, further validating the effectiveness of the GA-SVM-SVM:GA-SVM-SVM model for virtual screening. Table \ref{tab:virtual_screening_result} provides insights into the number of selected molecules from the BindingDB database for each target, based on different predetermined thresholds. Notably, this table reveals the presence of 4454, 803, 438, and 378 new inhibitor molecules for EGFR+HER2, ER, NF-$\kappa$B, and PR, respectively. These novel inhibitors were selected with 90\% precision in both the active/inactive and therapeutic classification decision-making processes.

\begin{figure*}[!ht]
    \centering
    \includegraphics[width=0.98\textwidth]{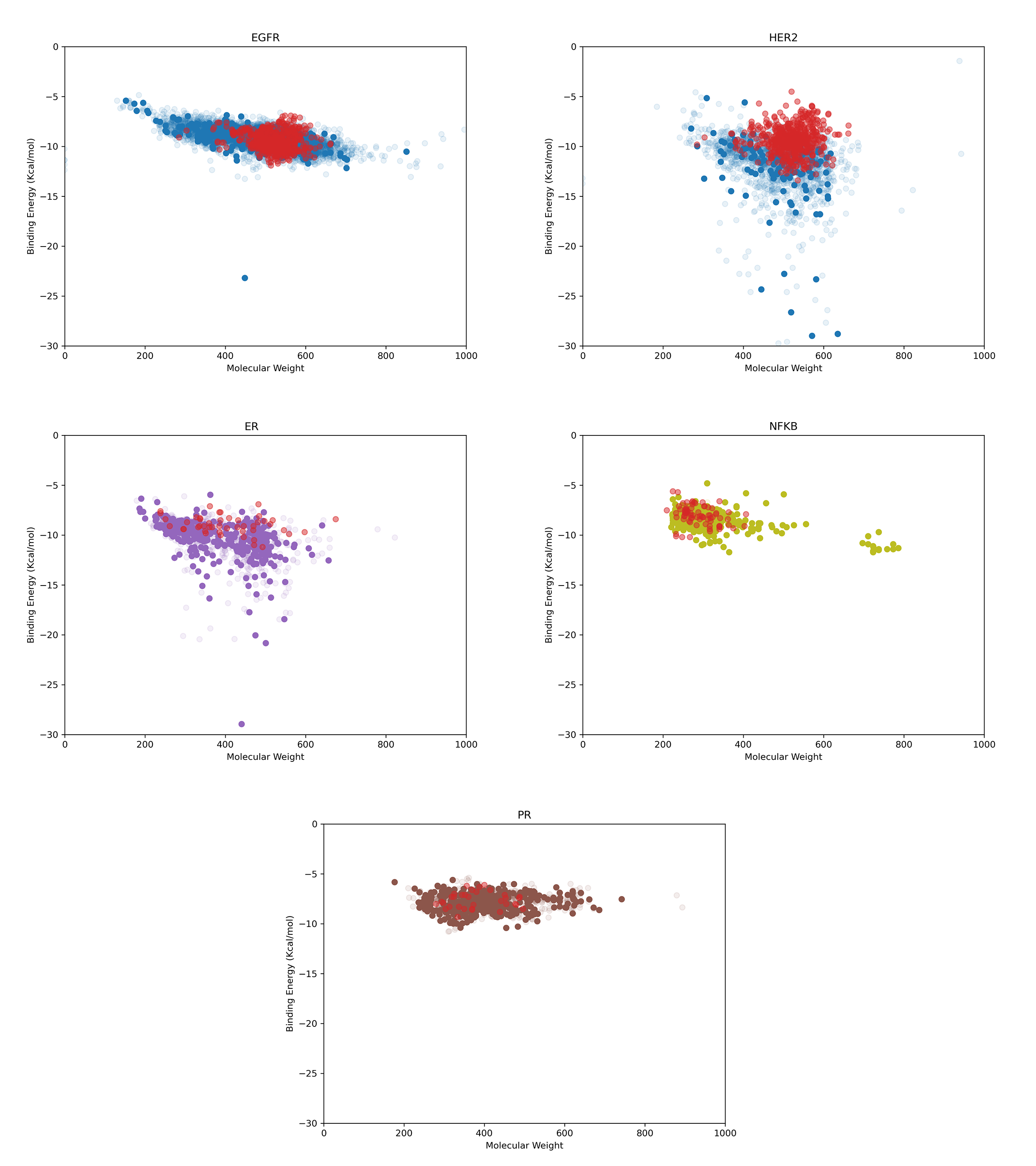}
    \caption{Docking results of new ligands obtained from virtual screening. The pale dots in the following plots represent the active molecules in the BindingDB database for each class, the filled dots represent the molecules that participated in the construction of the model, and the red dots are the new molecules proposed by the model obtained from the screening.}
    \label{fig:docking_result}
\end{figure*}

\begin{figure*}[!ht]
    \centering
    \includegraphics[width=0.6\textwidth]{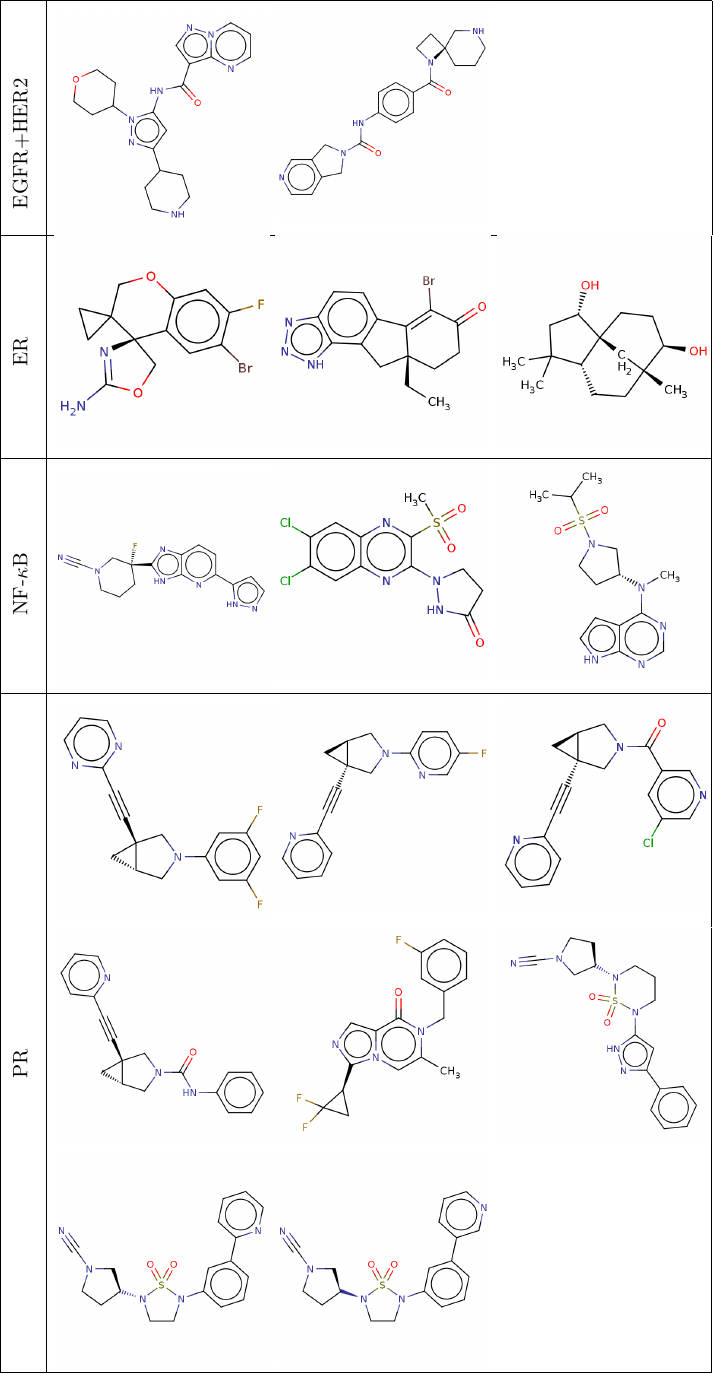}
    \caption{List of the molecules which are met all criteria.}
    \label{fig:molecules}
\end{figure*}

In order to easily identify the target for each molecule, our objective was to extract a straightforward rule from the model. To achieve this, we employed permutation importance to determine the significance of each feature in the model. Additionally, we utilized the Pearson method to create a hierarchical clustering dendrogram, which helped us identify the correlation distance (Euclidean distance) between features. Figure \ref{fig:features} displays the feature importance and hierarchical clustering dendrogram. Based on these findings, we constructed a simple questionnaire dendrogram for determining the target of each molecule which are selected with 90\% precision in both the active/inactive and target classification decision-making, as illustrated in Figure \ref{fig:questionnaire_dendrogram}. The data presented in Figure \ref{fig:questionnaire_dendrogram} provide concise and effective structure-activity relationship (SAR) information regarding the inhibitors. For instance, NF-$\kappa$B inhibitors exhibit significantly lower values for molecular walk count of order 10 (MWC10) and signal 10/weighted by polarizability (Mor10p) compared to other inhibitors. Additionally, EGFR and HER2 inhibitors demonstrate higher values of frequency of C–N at topological distance 3 (F03[C-N]) and SHED Donor-Donor (SHED\_DD) in comparison to ER and PR inhibitors. The ratio of eigenvalue n.1 from augmented edge adjacency mat. weighted by dipole moment (Eig01\_AEA(dm)), radial distribution function–105/weighted by polarizability (RDF105p), and radial distribution function–105/weighted by I-state (RDF105s) differentiates PR and ER inhibitors. These SAR information types effectively filter a significant portion of large databases, thus accelerating early-stage drug discovery projects that begin with extensive databases like GDB-13\cite{blum_970_2009}. The classification of molecules based on their therapeutic targets has garnered considerable attention in the field of chemoinformatics\cite{jalali-heravi_classification_2011}. These types of classifiers expand on the concept of ``Chemography''\cite{von_korff_assessing_2006,jalaliheravi_navigating_2012}, which refers to the art of navigating through chemical space. As evident from these figures, the inhibitors cluster within specific regions of the selected chemical space, aligning with the core objective of chemography.

In order to assess the binding energy of the molecules selected using the GA-SVM-SVM:GA-SVM-SVM model, we employed molecular docking for both the chosen molecules (with a multiplication of precision product exceeding 0.9 for both active/inactive and target classification) and the molecule sets within each class. The distribution of binding energy for these molecules, based on their molecular weights, is depicted in Figure \ref{fig:docking_result}. In these plots, the pale dots represent the active inhibitors labeled by the bindingDB database, while the filled dots represent the active molecules utilized in constructing the GA-SVM-SVM:GA-SVM-SVM model. The red dots correspond to new inhibitors, which exhibit binding energy within the range of -15 to -5 \si{\kilo\calorie\per\mole}. This range of binding energy proves to be sufficiently suitable for forming protein-ligand complexes.

In order to prioritize further study on the new molecules, we applied several medicinal criteria. The Lipinski rule suggests that ligands with a molecular weight of less than or equal to 500 ($M_W \leq 500$), a logarithm of the n-octanol/water distribution coefficient of less than or equal to 5 ($logP \leq 5$), a number of hydrogen bond acceptors of less than or equal to 10 ($H_{acc} \leq 10$), and a number of hydrogen bond donors of less than or equal to 5 ($H_{don} \leq 5$) exhibit good absorption or permeability. Accordingly, 376, 59, 91, and 35 ligands were accepted based on the Lipinski rule for EGFR+HER2, ER, NF-$\kappa$B, and PR targets, respectively. The Pfizer rule states that ligands with a $logP \geq 3$ and a topological polar surface area of less than or equal to 75 ($TPSA \leq 75$) are likely to be toxic. Consequently, 577, 30, 38, and 15 ligands passed the Pfizer rule for EGFR+HER2, ER, NF-$\kappa$B, and PR targets, respectively. The GSK rule suggests that ligands with a $MW \leq 400$ and $logP \leq 4$ may have a more favorable ADMET (absorption, distribution, metabolism, excretion, and toxicity) profile. Thus, 6, 11, 65, and 12 ligands were accepted based on the GSK rule for EGFR+HER2, ER, NF-$\kappa$B, and PR targets, respectively. Additionally, the golden triangle hypothesis proposes that ligands with a $200 \leq MW \leq 500$ and a logD (logarithm of the n-octanol/water distribution coefficient at pH = 7.4) ranging from -2 to 5 ($-2 \leq logD \leq 5$) may have a more favorable ADMET profile. Consequently, 166, 53, 91, and 34 ligands fulfilled the golden triangle criteria for EGFR+HER2, ER, NF-$\kappa$B, and PR targets, respectively. Moreover, several parameters such as QED (desirability functions based on eight drug-likeness related properties including $MW$, $log P$, $N_{HBA}$, $N_{HBD}$, $PSA$, $N_{rotb}$, $N_{Ar}$), SAscore (synthetic accessibility score based on a combination of fragment contributions and a complexity penalty), and MCE-18 (medicinal chemistry evolution in 2018 score molecules by novelty in terms of their cumulative sp3 complexity) were considered favorable in the medical industry. Ligands with QED greater than 0.67, SAscore less than 6, and MCE-18 larger than 45 were deemed desirable. Accordingly, 6, 14, 4, and 14 ligands met these criteria for EGFR+HER2, ER, NF-$\kappa$B, and PR targets, respectively. The distributions of selected molecules according to these rules are illustrated in Figures S3-S7. The molecules depicted in Figure \ref{fig:molecules} satisfy all these criteria. Specifically, 2, 3, 3, and 8 ligands met all the parameters for EGFR+HER2, ER, NF-$\kappa$B, and PR targets, respectively. As observed in this figure, each ligand within each class exhibits unique structural properties.

\section{Conclusion}
In this research, we utilized various machine learning methods and employed the GA-SVM-SVM and GA-RF-RF models for the active/inactive classification, as well as GA-SVM-SVM and GA-QDA-SVM for therapeutic classification. Based on the revenue generated by these methods in the pipeline, we selected GA-SVM-SVM:GA-SVM-SVM as the most effective pipeline for virtual screening. This model screened the BindingDB database inhibitors with varying precision ratios. As a result, we identified 4454, 803, 438, and 378 new inhibitor molecules for EGFR+HER2, ER, NF-$\kappa$B, and PR, respectively. Moreover, we introduced a simple dendrogram to determine the target of each inhibitor with new ligands which are predicted with over 90\% precision in both active/inactive and target prediction.

The molecules that exhibited a precision product exceeding 0.9 for both active/inactive and target classification were selected for further evaluation of their binding energy. Molecular docking analysis revealed that the binding energies of these inhibitors to their therapeutic targets ranged from -15 to -5 \si{\kilo\calorie\per\mole}, which is considered suitable for inhibiting the targets.

To prioritize further investigation of these new molecules, we applied several filters, including the Lipinski, Pfizer, GSK, golden triangle rules, as well as QED, SAscore, and MCE-18 parameters. Among these filters, 2, 3, 3, and 8 ligands met all the specified criteria for EGFR+HER2, ER, NF-$\kappa$B, and PR targets, respectively. We believe that this study can provide valuable insights for researchers working on the discovery of new inhibitors for breast cancer.

\begin{acknowledgement}

The authors acknowledge the UNESCO UNISA iThemba-LABS/NRF Africa Chair in Nanoscience \& Nanotechnology (U2ACN2) and the Centre for High-Performance Computing (CHPC), South Africa for providing computational resources and facilities for this research project. The authors would like to express their gratitude to all the members of the bioinformatics lab at Tarbiat Modares University (TMU) for their valuable contributions in the form of discussions and critical feedback on the manuscript.

\end{acknowledgement}

\begin{suppinfo}

Figures of sequence and structural alignment of EGFR and HER2 receptors, distribution of ligands according to the Lipinski, Pfizer, GSK, golden triangle rules and medical synthesis parameters and table of selected features for active/inactive and target are presented in supporting information.

\end{suppinfo}

\section{Data Availability}
Scripts of this study is available on Github, allowing users to access the source code. Visit our GitHub repository at https://github.com/parham-rezaee/bindingdb-bc-virtual-screening.

\end{document}